\documentclass[twocolumn]{jjap2} 
\title{Hall effect of quasi-hole gas in organic single-crystal transistors}

\author{Jun \textsc{Takeya}$^{1,2,3}$
\thanks{E-mail address: takeya@criepi.denken.or.jp},
Kazuhito \textsc{Tsukagoshi}$^{2,4}$,
Yoshinobu \textsc{Aoyagi}$^{2,5}$,
Taishi \textsc{Takenobu}$^{3,6}$
 and Yoshihiro \textsc{Iwasa}$^{3,6}$}

\inst{$^{1}$Materials Science Research Laboratory, CRIEPI, Tokyo 201-8511, Japan \\
$^{2}$RIKEN, Hirosawa 2-1, Wako 351-0198, Japan \\
$^{3}$Institute for Material Reasearch, Tohoku University, Sendai 980-8577, Japan \\
$^{4}$PRESTO, Japan Science and Technology Corporation, Kawaguchi 333-0012, Japan \\
$^{5}$Tokyo Institute of Technology, Yokohama 336-8502, Japan \\
$^{6}$CREST, Japan Science and Technology Corporation, Kawaguchi 333-0012, Japan
}

\abst{
Hall effect is detected in organic
field-effect transistors,
using appropriately shaped
rubrene (C$_{42}$H$_{28}$) single crystals.
It turned out that inverse Hall coefficient,
having a positive sign, is close to the amount
of electric-field induced charge upon the hole accumulation.
The presence of the normal Hall effect means
that the electromagnetic character of the surface charge
is not of hopping carriers but resembles that of
a two-dimensional hole-gas system.}

\kword{organic field-effect transistor, OFET, Hall effect, rubrene,
single-crystal FET}

\begin{document}
\maketitle


In quest of next-generation materials for fundamental
electronic components, organic semiconductors are
attractive due to their simple fabrication processes,
low production cost, capability of low energy synthesis,
and mechanical flexibility as well as sensitivity to light
\cite{dimitrakopoulos,podzorov_prl2}.
Although the organic semiconductors have been proven
to be available for the elemental circuit components,
represented by field-effect transistors (FETs),
the full performance of the materials would be
significantly reduced in broadly studied polymeric or
polycrystalline thin-film devices by extrinsic effects
such as those caused by grain boundaries \cite{horowitz}.
Indeed, recently developed single-crystal organic FETs (OFETs)
have demonstrated that the intrinsic FET performances
are superior to those reported for thin-film OFETs;
the FET mobility $\mu_{FET}$ reaches as high as 20 cm$^2$/Vs,
and the subthreshold swing is comparable to
that of normal single-crystal silicon FETs for an
aromatic molecular compound, rubrene \cite{menard,takeya2,podzorov_prl,
podzorov_apl,boer,takeya}.
The fact that a molecularly flat surface free from
dangling bonds is easily grown for the organic single
crystals \cite{takeya3}, may be partially responsible for the
apparent improvement in the device parameters.
Microscopic transport mechanism, however, is not yet
clear of the field-induced charge at the crystalline surface.
To find out the intrinsic potentials of these organic materials
and to foresee further applications of organic devices,
extended study of single-crystal OFETs is needed.

Historically, it has been debated for a long time
whether poorly doped aromatic organic semiconductors,
having the $\pi$ orbitals of adjacent molecules narrowly overlapped,
can realize band transport which justifies the ideal
electron-gas picture \cite{pope}.
Regarding the Hall effect, the reported results were
controversial even for bulk crystals in terms of sign,
temperature dependence and the value, possibly because
of difficulty in the measurement of poorly conductive
low-carrier-density systems \cite{pope}.
Note that the popular
expression of Hall coefficient $R_H \sim 1/Q$, where $Q$ denotes
the charge amount, is realized only when the \lq\lq charge cloud"
is spatially extended to lead to a good conductivity.
If the charge is basically localized, on the other hand,
$R_H$ would be much smaller, as is demonstrated for
amorphous semiconductors \cite{lecomber,friedman}.
In our present study, we use the single-crystal OFETs
so that the gate electric field moderately dopes charge
in the organic crystalline surface.
Detection of sizable Hall effect unambiguously tells us
that the charge cloud is well extended.

In our experiments, we crystallized rubrene molecules to
a thin platelet with a thickness of approximately 1 $\mu$m
by physical vapor transport.
To fabricate an FET,
the thin rubrene crystal is laminated on a SiO$_2$ / doped Si
substrate with gold electrodes patterned for both transverse
and longitudinal voltage detections;
starting from a SiO$_2$/doped Si wafer (SICO, GmbH) with a 500-nm-thick thermally
oxidized surface,
the SiO$_2$ surface is spin coated with HMDS (hexamethildisilane)
and six gold electrodes were deposited
at a thickness of 15 nm by vacuum evaporation through a shadow mask.
Figure 1 shows the top view of the device together with
an illustration of the measurement configuration.
As the transverse signal is only 0.1\% of the source-drain
voltage $V_D$ applied in the longitudinal direction,
it is essential to regulate the current and symmetrically
position the two narrow paths
from the gold electrodes at both sides so that the
potentials $V_1$ and $V_3$ at the opposite electrodes
are well balanced at zero magnetic field.
We trimmed the laminated rubrene crystal into a
Hall-bar shape using a scanning laser-etching equipment
for this purpose \cite{yagi}.
Such a dry-etching process
is preferable not to damage the device channel.



Four source measure units (SMUs) equipped in an Agilent
Technology E5207 semiconductor parameter analyzer were
used for the longitudinal and transverse measurements;
one applied the source-drain voltage and measured the
drain current $I_D$, another applied the gate voltage $V_G$,
and the other two detected voltage signals in either
transverse or longitudinal directions in the high-impedance
voltage-measurement mode.
Note that the input impedance
of the voltmeter is around T$\Omega$, which is high enough
for our measurement typically in a M$\Omega$ range.
The voltage signals are independent of the potential
drop at current injecting contacts, so that purely
intrinsic signals from the sample are measured while
avoiding contact resistances.
Magnetic field was applied perpendicular to the crystalline
surface using PPMS (Physical Properties Measurement System,
Quantum Design, Inc.).
The transverse voltage signals were
continuously measured at a sweeping gate voltage
from -30 to +30 V in 15 s and sweeping magnetic field repeatedly
in the range of +/-10 T with 0.27 T/min.

To detect the Hall effect, the transverse voltage $V^{trans}$
(=$V_3 - V_1$) is monitored during the continuous sweep of $V_G$,
under a slowly changing magnetic field, $H$, perpendicular
to the crystal surface.
With repeated sweep of $H$, $V^{trans}$
gradually changes with time.
Plotted in Fig. 2 are the
time evolutions of the transverse voltage for several gate
voltages, where a slow-drift component of $\sim$ 10 mV is already subtracted.
The plot apparently shows that $V^{trans}$ changes concomitantly
with the applied magnetic field and has maximum and minimum
values at +10 T and -10 T, respectively, which demonstrates
the presence of the Hall effect in OFETs.
The Hall coefficient $R_H$ is evaluated as $V^{trans} /\mu_0 H I_D$,
where $\mu_0$ denotes the permeability in vacuum, and inverse
$R_H$ is plotted as a function of $V_G$ in the upper panel of Fig. 3.
The sign of $R_H$ is positive for all negative $V_G$
(positive $H$ directs as defined in Fig. 1),
exhibiting a normal Hall effect of the holes induced by
the gate electric field.

For comparison,
 longitudinal conductance $\sigma$ are measured also
as functions of the gate voltage $V_G$.
$\sigma$ is evaluated
as $I_D / (V_2-V_1) L / W$ ($L$ and $W$ denote the length and
width of the measured portion of the channel),
and is plotted together with $I_D$ in Fig. 3.
The result well reproduces the standard model of FETs
for hole injection, i.e., $\sigma = -C_i(V_G-V_{th}) \mu_{FET}$,
where $C_i$ and $V_{th}$ denote the capacitance of the gate insulator
and the threshold voltage, respectively.
$\mu_{FET}$ can be evaluated as $\sim$ 1.5 cm$^2$/Vs, which is not
as high as the best value (20 cm$^2$/Vs) reported for
the air-gapped single-crystal FET measured in the most
conductive direction \cite{menard}.
Since our thin crystal is placed
in the least conductive direction (the $a$-axis direction)
for the measurement,
the anisotropy is at least partially responsible for
the discrepancy in $\mu_{FET}$.
The transfer characteristics
also show considerable positive threshold voltage,
indicating additional hole doping at the crystalline
surface either by treated SiO$_2$ surface or natural dopants
such as oxygen in the bulk crystal \cite{takeya2}.

We detect the Hall voltage only
when the crystalline surface has a good conductivity when $V_G < 0$.
As does the longitudinal conductivity $\sigma$
plotted together in the upper panel of Fig. 3, $1/R_H$ monotonically
increases with $|V_G-V_{th}|$.
Moreover, within 30\% of the magnitude, $1/R_H$ agrees
with $C_i |V_G-V_{th}|$, which is assumed to be the induced
surface charge upon the standard model of FETs,
as viewed also in Fig. 3.  Besides the 30-\% discrepancy,
the result corresponds to the free-electron model where
the charge amount $Q$ is equal to $1/R_H$. The result indicates
that a major part of the surface charge is highly mobile
so that its electromagnetic character resembles
that of extended holes.

To understand the observation in more detail, it is
convenient to evaluate the Hall mobility $\mu_H$ by calculating
the product of $R_H$ and $\sigma$, which are independently obtained in
our present experiments.  It is known that the ratio to the
longitudinal drift mobility $\mu_H/\mu_D$ ( = $R_H Q$) has a value
between 1 and 2 for band transport, whereas $\mu_H$ is much
smaller than $\mu_D$ ($1/R_H$ much larger than $Q$) when hopping
transport is dominant, because magnetic field does not
provide a transverse electromotive force for a single
tunnelling (hopping) process in principle.
As shown in the lower panel of Fig. 3, $\mu_H$ is in the range
from 1 to 1.5 cm$^2$/Vs depending on $V_G$.
The higher mobility
at a lower gate voltage may be attributed to the vertical
distribution of the field-induced charge further extending
deep into the crystal, where complication of the very surface
is smaller.
Although further study is required for the microscopic
understanding of the $\mu_H$ ($V_G$) profile,
the overall feature shows that
$\mu_H$ is more than 70\% of $\mu_{FET}$.
Since $\mu_{FET}$ can represent $\mu_D$, the above results
demonstrate the difficulty in assuming hopping transport
in the present case.
Although it has been argued that
a small Hall mobility can emerge when the interference among
different hopping processes contributes, this interpretation
is not favorable in the present case; $R_H$ should always be
negative in the interferential-hopping model regardless
of the sign of the responsible charge and $\mu_H/\mu_D$ does not
exceed 0.1 according to theoretical and experimental studies \cite{lecomber,friedman}.
In addition, as shown in the inset of Fig. 3, $\mu_{FET}$ is nearly
temperature-independent down to 260 K, which does not
resemble $\mu_{FET}(T)$ profile of hopping transport, either.

The above results indicate consistency in the band transport;
however, still the mobility of 1.5 cm$^2$/Vs may be small for
an ideal coherent transport for which truly extended wave
functions of electrons are responsible.
To have an idea of
how far the surface holes move, we estimate the mean free path
at room temperature ($T = 300$ K); the Boltzmann distribution
gives average velocity as
$\bar{v} = \sqrt{2 k_B T / m^*} \sim 1.1 \times 10^5$ m/s for
two-dimensional systems when the effective mass $m^*$ is nearly
equal to that of a free electron.
Giving the relaxation time
$\tau$ as $\tau = m^* \mu_{FET} / e \sim 10-15$ s, where $e$ denotes
the electron charge, the mean free path
$\ell$ ($= \bar{v} \tau \sim 0.11$ nm) can be estimated to be half of the lattice
constant $a$, showing apparent difference from typical band
metals in which $\ell \gg a$.
Even considering the presence of
high-energy particles and assuming mass enhancement by
polaronic renormalization \cite{podzorov_prl}, $\ell$ would not be much longer
than the lattice constant.
Therefore, although the clear
observation of a normal Hall effect suggests the extended
nature of the surface holes, they appear to be on the verge
of localization.

For future studies, the field-effect transistors may be a
suitable device to investigate how the extended electronic state is achieved
when the carriers are doped into the band insulator, as the carrier density
is continuously controlled by the gate electric field.
The two-dimensional electronic system around the crossover
between the extended and localized states is in itself intriguing,
analogous to the spontaneous construction of a large-scale network,
which is of general interest in the fields of physics,
statistics and social science \cite{watts}.
For technologies, further
understanding of the character of the holes responsible for
the FET action can lead to prescriptions to design
higher-performance OFETs, which would pave the way to low-power
applications such as logic-circuit components.
A device designed to utilize a channel deeper in a crystal,
for example, would be useful to provide higher $\mu_{FET}$ as
speculated above.  In these aspects,
the Hall-effect measurement of OFETs is expected to be a
powerful tool in the study of microscopic transport mechanisms,
as has been successfully performed for various electronic systems.

The authors thank I. Tsukada and T. Kuroda for their
technical assistance.
This work is partially supported by a Grant-in-Aid
for Scientific Research (No. 16740214) from the
Ministry of Education, Culture, Sports, Science,
and Technology, Japan.



\begin{figure}[htbp]
\vspace{-3cm}
\includegraphics[width=8cm]{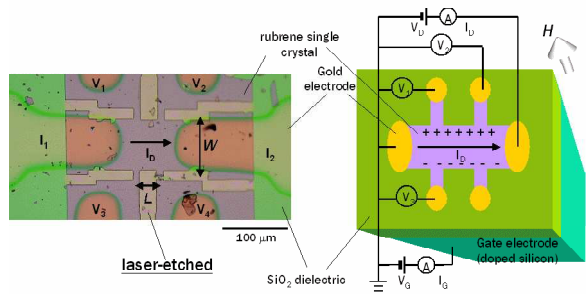}
\vspace{-3cm}
\caption{Schematic and optical view of rubrene single-crystal
field-effect transistor for Hall-effect measurement.
The left-hand side shows a top view of the sample that is used
for the detection of the Hall effect.
Dimensions of $L$ and $W$ are
$\sim 30$~$\mu$m and $\sim 100$~$\mu$m, respectively.
The configuration of the longitudinal and transverse voltage
measurements is illustrated on the right.
The direction of positive magnetic field is also defined.}
\label{f1}
\end{figure}



\begin{figure}[htbp]
\vspace{-3cm}
\includegraphics[width=8cm]{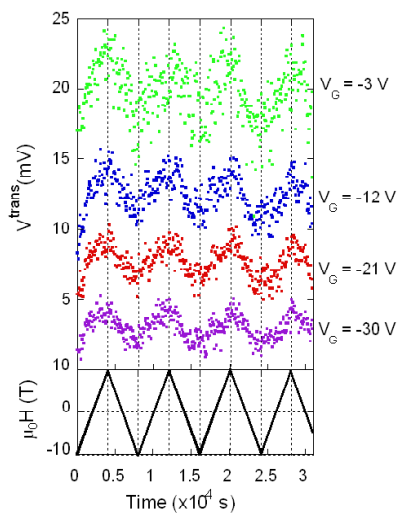}
\vspace{-1cm}
\caption{Hall signals of rubrene single-crystal field-effect
transistor under slowly varying magnetic field.
The transverse voltage signal $V^{trans}$ (= $V_1-V_3$) is plotted as
a function of time for several gate voltages under a magnetic
field repeatedly swept as shown at the bottom.
}
\label{f3}
\end{figure}


\begin{figure}[htbp]
\vspace{-2cm}
\includegraphics[width=8cm]{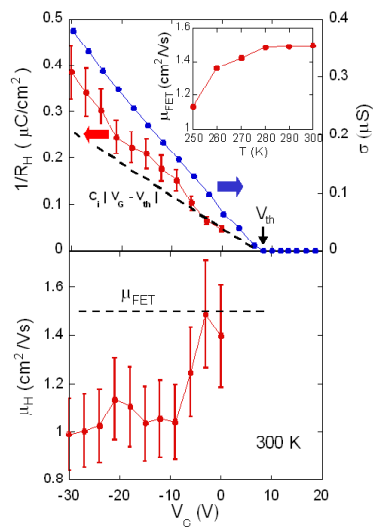}
\vspace{-2cm}
\caption{Inverse Hall coefficient and Hall mobility of
rubrene single-crystal field-effect transistor as function
of gate voltage.
Plotted in the upper panel are the inverse Hall coefficient
(red circles), longitudinal surface conductivity
(blue circles) and electric-field induced charge
(the black broken line) assumed in a standard model of
field-effect transistors.
Combining the results of both transverse and longitudinal
measurements, we estimate the Hall mobility $\mu_H$ shown in
the lower panel.
The inset shows the FET mobility $\mu_{FET}$ as a function
of temperature above 250 K.}
\label{f4}
\end{figure}


\end{document}